\documentclass[aps,prd,onecolumn,superscriptaddress,showpacs]{revtex4}

\usepackage{graphicx}

\begin{document}

\title{Bounds on $\gamma$ from CP violation measurements in
$B \rightarrow \pi^+ \pi^-$ and $B \rightarrow \psi K_S$}

\author{F.\ J.\ Botella}
\affiliation{Centro de F\'{\i}sica das Interac\c{c}\~{o}es Fundamentais,
	Instituto Superior T\'{e}cnico,
	P-1049-001 Lisboa, Portugal}
\affiliation{Departament de F\'{\i}sica Te\`{o}rica and IFIC,
        Universitat de Val\`{e}ncia-CSIC,
        E-46100, Burjassot, Spain}
\author{Jo\~{a}o P.\ Silva}
\affiliation{Centro de F\'{\i}sica das Interac\c{c}\~{o}es Fundamentais,
	Instituto Superior T\'{e}cnico,
	P-1049-001 Lisboa, Portugal}
\affiliation{Instituto Superior de Engenharia de Lisboa,
	Rua Conselheiro Em\'{\i}dio Navarro,
	1900 Lisboa, Portugal}

\date{\today}

\begin{abstract}
We study the determination of $\gamma$ from 
CP-violating observables in
$B \rightarrow \pi^+ \pi^-$ and $B \rightarrow \psi K_S$.
This determination requires theoretical input to one
combination of hadronic parameters.
We show that a mild assumption about this quantity may allow
bounds to be placed on $\gamma$,
but we stress the pernicious effects that an eightfold
discrete ambiguity has on such an analysis.
The bounds are discussed as a function of the direct ($C$) and
interference ($S$) CP-violating observables obtained
from time-dependent $B \rightarrow \pi^+ \pi^-$ decays,
and their behavior in the presence of new physics effects
in $B - \overline{B}$ mixing is studied.
\end{abstract}


\maketitle

\section{\label{sec:intro}Introduction}

The Standard Model (SM) of electroweak interactions has been 
so successful that increasingly detailed probes are required
in order to uncover possible new physics effects.
CP violation seems to provide a particularly promising probe,
because it appears in the SM through one single irremovable
phase in the Cabibbo-Kobayashi-Maskawa (CKM) matrix \cite{CKM}.
As a result,
the measurements of any two CP-violating experiments must be
related through CP-conserving quantities.
In principle,
this makes the SM a very predictive theory of CP violation.
In practice,
however,
the CP-conserving quantities required to extract weak interaction
parameters from experiment usually involve the strong interaction,
are difficult to calculate,
and the interpretations of the experiments in terms of
parameters of the original Lagrangian are plagued by hadronic
uncertainties.

One notable exception occurs with the determination of
$\sin{(2 \tilde{\beta})}$ from the time dependent
asymmetry in $B \rightarrow \psi K_S$.
In the SM,
and with the usual phase convention,
$\tilde{\beta} \equiv \beta$ is the phase of $V_{td}^\ast$.
The current world average is \cite{Browder}
\begin{equation}
\sin{(2 \tilde{\beta})} = 0.736 \pm 0.049,
\end{equation}
based on the very precise measurements by
Babar \cite{Babar} and Belle \cite{Belle}.
In this article,
we will also consider the possibility that there might be
new physics contributions affecting the phase of
$B - \overline{B}$ mixing \cite{note-magnitude-BBbar}.
In that case,
the phase $\tilde{\beta}$ measured in $B \rightarrow \psi K_S$
decays does \textit{not} coincide with the phase $\beta$ of $V_{td}^\ast$.

The current measurement of $\beta$ is in
agreement with measurements based on $|V_{ub}/V_{cb}|$,
$\Delta m_B$, $\Delta m_{Bs}$,
and CP violation in the $K -\overline{K}$ system,
although each of these is plagued by hadronic uncertainties.

So,
we would like to constrain the CKM source of CP violation
in as many ways as possible,
in the hope of uncovering new physics effects.
One possibility arises in the time-dependent asymmetry
in $B \rightarrow \pi^+ \pi^-$ decays.
If there were only contributions from tree level diagrams,
this would provide a clean measurement of
$\sin{(2 \tilde{\beta} + 2 \gamma)}$ \cite{BLS}.
Unfortunately,
the presence of penguin contributions with a different weak
phase imply that this measurement is plagued by hadronic uncertainties.
One way out of this problem consists in estimating this penguin
pollution in some way (see section~\ref{sec:exptheor}).
Recently,
Buchalla and Safir (BS) have proposed a different approach \cite{BS}.
In their method,
a mild assumption is made about the needed theoretical input
in order to derive a bound on $\gamma$,
which is valid provided the interference CP violation observable
in $B \rightarrow \pi^+ \pi^-$ (S) lies above $- \sin{(2 \tilde{\beta})}$.
Their bound holds within the SM
(although this is not obvious from their article)
and  can be obtained in the limit of no penguin pollution that
corresponds to setting the direct CP violation observable
in $B \rightarrow \pi^+ \pi^-$ (C) to zero.

In this article,
we extend their result by studying which bounds occur
when $C \neq 0$ and when there are new physics in 
$B - \overline{B}$ mixing (that is, when $\tilde{\beta} \neq \beta$).
Because we do not go through the Wolfenstein's  parameters
$\rho$ and $\eta$ \cite{Wolf},
we obtain, as a particular case,
an easier derivation of the BS result.
Our analysis will allow us to state what types of new physics
effects are subject to these new bounds and how they change
from $C=0$ to $C \neq 0$.
In particular,
we will stress the very important impact that an eightfold
discrete ambiguity has on such bounds.

Although that is not the main point of this article,
we will also comment briefly on how one can perform the
precise determination of $\gamma$,
once a specific value for the relevant hadronic quantity is taken.
In doing so,
we differ from previous analysis in that we start from the
experimental values for $C$ and $S$
(using only the \textit{one} piece of theoretical input required)
rather than plot ``predictions'' for $C$ and $S$ given a number
of theoretical inputs.
The end result is the same, but this is closer in spirit to what
one really wishes to do; we wish to use experimental results in order
to learn about theoretical parameters and not the converse.

Our article is organized as follows.
In section~\ref{sec:exptheor},
we set the notation,
introducing the relevant experimental and
theoretical quantities involved in $B \rightarrow \pi^+ \pi^-$
decays.
In section~\ref{sec:master} we develop the two formulas which will
guide our analysis of the bounds on $\gamma$ discussed
in section~\ref{sec:bounds}.
We draw our conclusions in section~\ref{sec:conclusions} and include
a trivial but useful inequality in the appendix.

\section{\label{sec:exptheor}Experimental observables versus
Theoretical parameters}

The time-dependent CP asymmetry in $B \rightarrow \pi^+ \pi^-$ decays 
may be written as\cite{BLS}
\begin{eqnarray}
A_{CP}(t)
&=&
\frac{
\Gamma[B(t) \rightarrow \pi^+ \pi^-] - 
\Gamma[\overline{B}(t) \rightarrow \pi^+ \pi^-]}{
\Gamma[B(t) \rightarrow \pi^+ \pi^-] + 
\Gamma[\overline{B}(t) \rightarrow \pi^+ \pi^-]}
\nonumber\\
&=&
- S \sin{\Delta m_B\, t} + C \cos{\Delta m_B\, t},
\end{eqnarray}
where
\begin{equation}
S = \frac{2 {\textrm Im} \lambda}{1+|\lambda|^2}
\ \ \ \ \ \ \ \ 
C = \frac{1-|\lambda|^2}{1+|\lambda|^2}.
\label{eq:CandS}
\end{equation}
Clearly $S^2 + C^2 \leq 1$.
The corresponding experimental results are \cite{Babar-pipi,BELLE-pipi}
\begin{eqnarray}
S & = &
\left\{
\begin{array}{ll}
-0.40 \pm 0.22 \pm 0.03 & (\textrm{Babar})\\
-1.23 \pm 0.41^{+0.08}_{-0.07} & (\textrm{Belle})
\end{array}
\right.
\nonumber\\
C & = &
\left\{
\begin{array}{ll}
-0.19 \pm 0.19 \pm 0.05 & (\textrm{Babar})\\
-0.77 \pm 0.27 \pm 0.08 & (\textrm{Belle})
\end{array}
\right.
\label{eq:exp-results}
\end{eqnarray}
which the Heavy Flavour Averaging Group combines into
$S = -0.58 \pm 0.20$ and $C= -0.38 \pm 0.16$ \cite{HFAG}.
Eqs.~(\ref{eq:CandS}) and (\ref{eq:exp-results}) 
imply that $\lambda$ is a quantity accessible experimentally,
up to a twofold discrete ambiguity \cite{Dun95} in its real part:
\begin{equation}
\pm \sqrt{1 - C^2 - S^2}= \frac{2 {\textrm Re} \lambda}{1+|\lambda|^2}.
\label{eq:real-lambda}
\end{equation}

On the other hand, $\lambda$ may be written in terms of
theoretical parameters as
\begin{equation}
\lambda = \frac{q}{p} \frac{\bar A}{A},
\label{lambda-general}
\end{equation}
where $q/p$ reflects the $B - \overline{B}$ mixing,
and $A$ ($\bar A$) is the amplitude for the
$B \rightarrow \pi^+ \pi^-$ ($\overline{B} \rightarrow \pi^+ \pi^-$)
decay.
With the usual phase convention for the CKM parameters,
these quantities may be written in terms of weak and strong interaction
parameters as
\begin{eqnarray}
\frac{q}{p} 
& = &
e^{-2 i \tilde{\beta}},
\nonumber\\
A = V_{ub}^\ast  V_{ud}\, T + V_{cb}^\ast V_{cd}\, P\, e^{i \delta}
& = & 
|V_{ub} V_{ud}\, T| \left( e^{i \gamma} + z \right),
\nonumber\\
\bar A = V_{ub} V_{ud}^\ast\, T + V_{cb} V_{cd}^\ast P\, e^{i \delta}
& = & 
|V_{ub} V_{ud}\, T| \left( e^{-i \gamma} + z \right),
\label{parametrization}
\end{eqnarray}
where $P$ and $T$ are magnitudes of hadronic quantities,
$\delta$ is a strong phase difference,
and
\begin{equation}
z = \frac{P/T}{R_b} e^{i \delta} = r e^{i \delta}
\end{equation}
includes a dependence on the weak parameter
$R_b = |V_{ub} V_{ud}|/|V_{cb} V_{cd}| = \sqrt{\rho^2 + \eta^2}$.
The weak phase $\tilde{\beta}$ coincides with the CKM
parameter $\beta$,
if one stays within the framework of the SM;
but $\tilde{\beta}$ may differ from $\beta$,
if there are new physics contributions affecting the phase in 
$B - \overline{B}$ mixing \cite{note-magnitude-BBbar}.

Substituting Eqs.~(\ref{parametrization}) in
Eq.~(\ref{lambda-general}), we obtain
\begin{equation}
\lambda = e^{-2i \tilde{\beta}}\  \frac{e^{-i \gamma} + z}{e^{i \gamma} + z}.
\label{lambda-master}
\end{equation}
This equation relates the measurable quantity on the LHS with the
theoretical quantities on the RHS.
In some sense,
these two sides are usually kept apart.
(As we shall see in section~\ref{sec:master},
great simplifications occur if we subvert this standard practice.)
One may now substitute Eq.~(\ref{lambda-master}) in
Eqs.~(\ref{eq:CandS}) to find \cite{note2}
\begin{eqnarray}
S 
& = &
-
\frac{\sin{(2 \tilde{\beta}+ 2 \gamma)} + 
2 r \sin{(2 \tilde{\beta}+ \gamma)} \cos{\delta} +
r^2 \sin{(2 \tilde{\beta})} }{
1 + 2 r \cos{\gamma} \cos{\delta} + r^2},
\label{eq:S-BS}
\\
C
& = &
\frac{2 r \sin{\gamma} \sin{\delta} }{
1 + 2 r \cos{\gamma} \cos{\delta} + r^2},
\label{eq:C-from-theor}
\\
\frac{2 {\textrm Re} \lambda}{1+|\lambda|^2}
& = &
\frac{\cos{(2 \tilde{\beta}+ 2 \gamma)} + 
2 r \cos{(2 \tilde{\beta}+ \gamma)} \cos{\delta} +
r^2 \cos{(2 \tilde{\beta})} }{
1 + 2 r \cos{\gamma} \cos{\delta} + r^2}.
\end{eqnarray}
If there were no penguin amplitudes ($r=0$),
then $\lambda$ would be given by the pure phase
$ - 2 \tilde{\beta} - 2 \gamma$,
which is \textit{by definition} equal do
$2 \alpha$ (mod.\, $2 \pi$).
In that case, $C$ would vanish and $S$ would provide a clear
determination of the phase $\tilde{\beta} + \gamma$,
which in the SM coincides with $\beta + \gamma$.
As is well know, the presence of the ``penguin pollution''
spoils this determination.
In fact,
since $\tilde{\beta}$ has been determined in $B \rightarrow \psi K_S$
decays,
there are two experimental observables ($C$ and $S$)
and three unknowns
($r$, $\delta$, and the weak phase $\gamma$).
One needs some extra piece of information about the
hadronic parameters $r$ and $\delta$ in order to determine the
weak phase $\gamma$.

This extra information may be achieved in a variety of ways.
Gronau and London \cite{GL} used isospin to relate
$B \rightarrow \pi^+ \pi^-, \pi^0 \pi^0$ and
$B^+ \rightarrow \pi^+ \pi^0$ decays.
Their method has received renewed life from the recent announcements
by Babar \cite{Babar-pi0pi0} and Belle \cite{BELLE-pi0pi0}
of a large branching ratio for the $\pi^0 \pi^0$ final state.
Silva and Wolfenstein \cite{SW94} proposed an estimate of the
penguin contribution through an SU(3) relation between
 $B \rightarrow \pi^+ \pi^-$ and $B \rightarrow K^+ \pi^-$.
Chiang, Gronau and Rosner \cite{CGR} used SU(3) to estimate P/T 
from a variety of observables.
Alternatively,
one may estimate $r$ and $\delta$ directly from theory,
within QCD factorization \cite{BBNS};
Buchalla and Safir quote
$r R_b = 0.107 \pm 0.031$ and $\delta = 0.15 \pm 0.25$ \cite{BS}.

One could try to proceed without the extra piece of information.
Working within the SM,
one could substitute
$\tilde{\beta} = \beta$ and $\gamma$ by $\rho$ and $\eta$
on the RHS of Eq.~(\ref{eq:S-BS}),
which would lead to a rather complicated expression.
(Notice that this substitution is only possible within
the SM, since, in general, $\tilde{\beta}$ is not related
to $\rho$ and $\eta$.)
Such work has been done recently by Buchalla and Safir \cite{BS},
who point out that a lower bound on $\eta$ (and, thus, $\gamma$)
can be achieved with a mild assumption on the hadronic parameters,
as long as $S > - \sin{(2 \beta)}$.
The value of that lower bound is equal to the value that one would
obtain for $\eta$ ($\gamma$) in the limit of vanishing
penguin amplitude (that is, with C=0).
In what follows,
we will recover their result in a much simpler way,
by evading any mention of $\rho$ and $\eta$.
This will allow us to generalize their result
for $C \neq 0$ and to discuss how such bounds 
are affected by possible new physics contributions
to the $B - \overline{B}$ mixing phase.

\section{\label{sec:master}Two master formulae}

We start from Eq.~(\ref{lambda-master}),
and multiply both sides by the denominator of the RHS.
Reordering the terms, we obtain
\begin{equation}
\lambda e^{i \gamma} - e^{- 2 i \tilde{\beta}} e^{-i \gamma}
=
z (e^{- 2 i \tilde{\beta}} - \lambda). 
\end{equation}
Equating the real and imaginary parts
\begin{eqnarray}
\textrm{Re}(\lambda - e^{-2 i \tilde{\beta}}) \cos{\gamma}
- 
\textrm{Im}(\lambda + e^{-2 i \tilde{\beta}}) \sin{\gamma}
& = &
\textrm{Re}\left[ z (e^{-2 i \tilde{\beta}}- \lambda) \right],
\nonumber\\
\textrm{Im}(\lambda - e^{-2 i \tilde{\beta}}) \cos{\gamma}
+
\textrm{Re}(\lambda + e^{-2 i \tilde{\beta}}) \sin{\gamma}
& = &
\textrm{Im}\left[ z (e^{-2 i \tilde{\beta}}- \lambda) \right],
\end{eqnarray}
we find
\begin{eqnarray}
\cos{\gamma} 
& = &
- \textrm{Re}(z)
- \frac{2\,\textrm{Im}(\lambda e^{2 i \tilde{\beta}})}{1-|\lambda|^2} 
\, \textrm{Im}(z),
\nonumber\\
\sin{\gamma}
& = &
\frac{
(1+|\lambda|^2) - 2\, \textrm{Re}(\lambda e^{2 i \tilde{\beta}})
}{1-|\lambda|^2}
\, \textrm{Im}(z).
\label{eq:sin-cos-1}
\end{eqnarray}

Using Eqs.~(\ref{eq:CandS}) and (\ref{eq:real-lambda}),
we may rewrite Eqs.~(\ref{eq:sin-cos-1}) as
\begin{eqnarray}
\cos{\gamma_\pm} 
& = &
- \textrm{Re}(z)
- \frac{I_\pm}{C}\, \textrm{Im}(z),
\nonumber\\
\sin{\gamma_\pm}
& = &
\frac{1 - R_\pm}{C}\, \textrm{Im}(z),
\label{eq:sin-cos-2}
\end{eqnarray}
where
\begin{eqnarray}
R_\pm
& = &
\frac{2\,\textrm{Re}(\lambda e^{2 i \tilde{\beta}})}{1+|\lambda|^2} 
=
\pm \sqrt{1 - C^2 - S^2} \cos{(2 \tilde{\beta})}
- S \sin{(2 \tilde{\beta})},
\nonumber\\
I_\pm
& = &
\frac{2\,\textrm{Im}(\lambda e^{2 i \tilde{\beta}})}{1+|\lambda|^2} 
=
S \cos{(2 \tilde{\beta})}
\pm \sqrt{1 - C^2 - S^2} \sin{(2 \tilde{\beta})},
\label{eq:RandI}
\end{eqnarray}
are determined exclusively from experiment,
with the discrete ambiguity present in Eq.~(\ref{eq:real-lambda}).
It is easy to show (\textit{c.f.} the appendix) that
$|R_\pm|$ and $|I_\pm|$ are bounded by $\sqrt{1 - C^2}$.

Eqs.~(\ref{eq:sin-cos-2}) depend on two different combinations
of hadronic parameters, which we may choose as
$\{ \textrm{Re}(z), \textrm{Im}(z) \}$
or as
$\{ r, \delta\}$.
As we know from the parameter counting of the previous section,
one combination of hadronic parameters will always remain.
The other combination may be eliminated in a variety of ways.
For example,
\begin{equation}
\tan{\gamma_\pm}
=
\frac{-1 + R_\pm}{C \cot{\delta} + I_\pm},
\label{master-Joao}
\end{equation}
or
\begin{equation}
I_{\pm} \sin{\gamma_\pm} +
(1 - R_\pm) (\cos{\gamma_\pm} + Q) = 0,
\label{master-Quico}
\end{equation}
with
\begin{equation}
Q = r \cos{\delta}
\label{eq:Q}
\end{equation}
A few comments are in order.
First,
Eq.~(\ref{master-Joao}) has a form which will allow us to
derive a bound on $\gamma$ which generalizes the results of
BS in a very clear way.
Second,
for a given set of experimental values for
$\sin{(2 \tilde{\beta})}$, $C$, and $S$,
the theoretical parameter $\delta$ \textit{cannot} 
take an arbitrary value.
For example,
if $C \neq 0$ then $\delta$ cannot vanish,
as is easily seen from Eq.~(\ref{eq:C-from-theor}).
Third,
we have found numerically that,
even if one takes a value of $\delta$ consistent with the
experimental observables,
Eq.~(\ref{master-Joao}) is very sensitive to the exact value
chosen for $\delta$.
For the previous reasons,
and although Eq.~(\ref{master-Joao}) is so well suited
to study the bounds on $\gamma$,
Eq.~(\ref{master-Quico}) is more useful when studying the
dependence of $\gamma$ on the theoretical parameters (through $Q$).
Finally,
for a given set of experimental values of
$\sin{(2 \tilde{\beta})}$, $C$, and $S$,
and for the same $\cot{\delta}$,
there is an eightfold ambiguity in the determination of
$\gamma$.
A twofold ambiguity arises from the existence of
two values of $\tan{\gamma}$
($\tan{\gamma_+}$ and $\tan{\gamma_-}$)
which satisfy Eq.~(\ref{master-Joao}).
This is related to the ${\textrm Re} \lambda$ in
Eq.~(\ref{eq:real-lambda}), 
whose measurement would remove this twofold ambiguity,
and implies that $\lambda_+ = - \lambda_-^\ast$.
Of course, this is obtained for different values of $r$.
Another twofold ambiguity arises from the unknown
sign of $\cos{(2 \tilde{\beta})}$.
The compound transformation
$\cos{(2 \tilde{\beta})} \rightarrow -\cos{(2 \tilde{\beta})}$
and $\cot{\delta} \rightarrow - \cot{\delta}$
leads to $\tan{\gamma_\pm} \rightarrow - \tan{\gamma_\mp}$.
The final twofold ambiguity arises from the inversion of
the function $\tan{\gamma}$,
and corresponds to a symmetry $\gamma \rightarrow \pi + \gamma$.
If there is no new physics in the $K - \overline{K}$ system,
and if we trust the sign of the bag parameter,
then this ambiguity is removed since $\gamma$ cannot lie
outside $(0,\pi)$.

In the next section,
we will use Eqs.~(\ref{master-Joao}) and (\ref{master-Quico})
together with mild assumptions on the hadronic parameters
$\delta$ and $Q$ (respectively),
in order to provide model independent bounds on the CKM phase
$\gamma$.

\section{\label{sec:bounds}Bounds on $\gamma$}

\subsection{\label{subsec:bounds-on-tan+-}Bounds on $\tan{\gamma_\pm}$}

We will now study Eq.~(\ref{master-Joao}) in more detail.
Every quantity on the RHS of that equation is determined from
experiment, aside from $\cot{\delta}$.
It would be nice to be able to relate $\tan{\gamma_\pm}$ to
the value
\begin{equation}
L_\pm
=
\frac{-1 + R_\pm}{I_\pm}
=
\frac{-1\pm \sqrt{1 - C^2 - S^2} \cos{(2 \tilde{\beta})}
- S \sin{(2 \tilde{\beta})}}{S \cos{(2 \tilde{\beta})}
\pm \sqrt{1 - C^2 - S^2} \sin{(2 \tilde{\beta})}}
\label{eq:L+-}
\end{equation}
obtained from Eq.~(\ref{master-Joao}) by suppressing the
$C \cot{\delta}$ term.
If we knew, for example, that $C \cot{\delta}$ were positive,
then we might be able to derive a bound on $\tan{\gamma_\pm}$.
Notice that,
using Eq.~(\ref{eq:C-from-theor}),
$C \cot{\delta} \propto \cos{\delta}$.
Therefore,
within the current range for $\gamma$,
$C \cot{\delta} > 0$ as long as $-\pi/2 \leq \delta \leq \pi/2$,
which is a very mild constraint on $\delta$.
Indeed,
$\delta$ is expected to be small on general grounds;
BS quote $\delta = 0.15 \pm 0.25$ based on QCD factorization \cite{BS}.

To proceed we note that, because $|R_\pm| \leq \sqrt{1-C^2}$,
the numerator in Eq.~(\ref{master-Joao}) cannot be positive;
$-1+R_\pm \leq 0$.
As for the denominator,
if $C \cot{\delta} \geq 0$, then $C \cot{\delta} + I_\pm \geq I_\pm$.
But, the sign of this inequality upon invertion depends
on the sign of $(C \cot{\delta} + I_\pm)\, I_\pm$.
When all is taken into account, we obtain:
\begin{eqnarray}
\tan{\gamma_\pm} \geq L_\pm
\ \ \ \textrm{if}\ \ \ 
C \cot{\delta}\, (C \cot{\delta} + I_\pm)\, I_\pm > 0,
\label{eq:lower-bound}
\\
\tan{\gamma_\pm} \leq L_\pm
\ \ \ \textrm{if}\ \ \ 
C \cot{\delta}\, (C \cot{\delta} + I_\pm)\, I_\pm < 0.
\label{eq:upper-bound}
\end{eqnarray}
These equations generalize the bound of Buchalla and Safir
and constitute the main result of this article.
(We note that the $(C \cot{\delta} + I_\pm)\, I_\pm$ piece
of the conditions on the right-hand-side of these equations arise from
the $\arctan{x}$ function going through $\pm 90^\circ$,
and not from a change from an upper to a lower bound on $\gamma$.
This clear from  Eqs.~(\ref{master-Joao}) and (\ref{eq:L+-})
and will also become apparent from the figures in the next section.)

These bounds enclose many important features.
First,
they depend only on $\tilde{\beta}$ and \textit{not} on $\beta$.
This is obvious from Eq.~(\ref{lambda-master}),
but would be hidden if we were to use $\rho$ and $\eta$ in
our analysis,
as done in \cite{BS},
since it is $\beta$ (not $\tilde{\beta}$) which is related to those
Wolfenstein parameters.
Second,
the \textit{sign} of $\cos{\tilde{\beta}}$ enters crucialy into the
bounds through Eq.~(\ref{eq:L+-}),
and this sign cannot be determined from the usual experiments
with $B \rightarrow \psi K_S$ decays \cite{sign-of-cosbeta}.
Third,
because of the $\pm$ discrete ambiguity,
one must analyze what happens to both solutions;
$\tan{\gamma_+}$ and $\tan{\gamma_-}$. 
This will depend on the exact values for $I_\pm$.
Fourth,
it is in principle possible that $\gamma_+$
satisfies Eq.~(\ref{eq:upper-bound}),
while $\gamma_-$ satisfies Eq.~(\ref{eq:lower-bound}).

The third point has one crucial consequence.
Imagine that we have measured values for
$\sin{(2 \tilde{\beta})}$, $C$, and $S$,
and that we assume that $\cos{(2 \tilde{\beta})}$ is positive.
Imagine also that these conditions allow us to establish that
$\tan{\gamma_-} \geq L_-$.
This will still not provide us with an absolute lower bound
on $\gamma$,
unless we can ensure
(either because $\tan{\gamma_+} > \tan{\gamma_-}$
or because we have some theoretical reason to exclude
the possibility that $\gamma = \gamma_+$)
that $\gamma_+$ is not below $\gamma_-$.
The biggest problems occur when $\gamma_+ \sim \gamma_-$,
which, given that
\begin{equation}
\tan{\gamma_+} - \tan{\gamma_-}
\propto
\sqrt{1 - C^2 - S^2}\
\left[
S + \sin{(2 \tilde{\beta})} + C \cot{\delta} \cos{(2 \tilde{\beta})}
\right],
\label{eq:crossing}
\end{equation}
occurs when
$S = - \sin{(2 \tilde{\beta})} - C \cot{\delta} \cos{(2 \tilde{\beta})}$.
This will be clear from the figures in the next section.

\subsection{A simple example}

\begin{figure}[htb]
\includegraphics*[height=7cm]{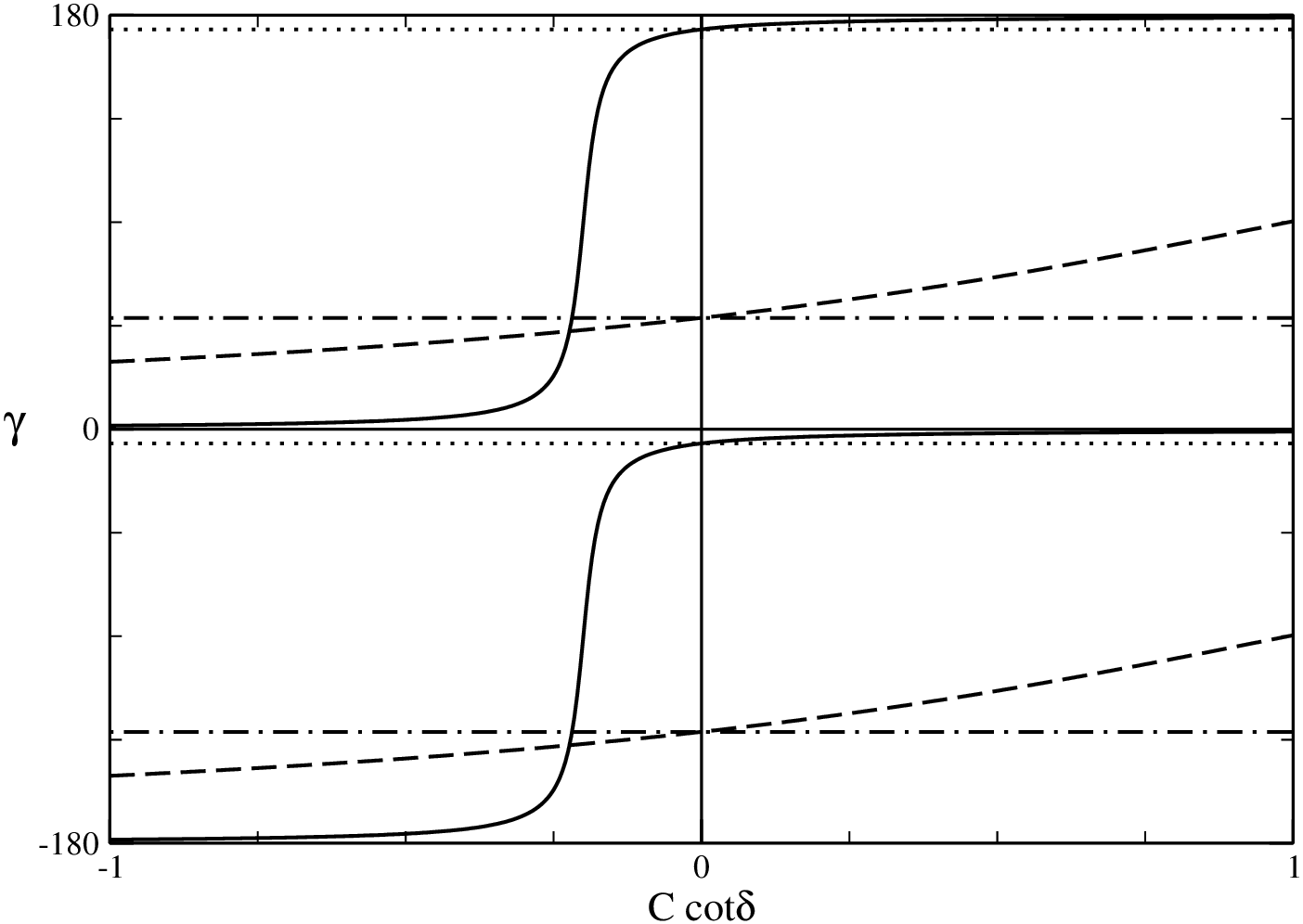}
\caption{\label{figure1}Values ``determined'' for $\gamma$,
as a function of the theoretical input for $C \cot{\delta}$,
assuming $\sin{(2 \tilde{\beta})}= 0.736$,
$C=0.0528$,
and $S=-0.585$.
The solid (dashed) curves correspond to $\gamma_+$ ($\gamma_-$).
The horizontal dotted (dash-dotted) lines correspond to
the values of $\gamma$ for which $\tan{\gamma}=L_+$
($\tan{\gamma}=L_-$).
Here we choose $\cos{(2 \tilde{\beta})}$ positive,
as in the SM.
}
\end{figure}

Let us consider
$\sin{(2 \tilde{\beta})}= 0.736$,
$C=0.0528$,
and $S=-0.585$.
(These putative experimental results can be ``fabricated'' with
the values $R_b=0.4$, $r\, R_b=0.11$, $\delta=0.15$, and $\gamma=60^\circ$.)
To start,
let us take the positive $\cos{(2 \tilde{\beta})} = + 0.677$.
The ``experimental'' observables become
$I_+=0.195$, $L_+ = -0.11$, $I_- = -0.992$ and $L_- = 1.12$.
In order to turn these experiments into a bound on $\gamma$,
we need some assumption about $C \cot{\delta}$.
Assuming that $C \cot{\delta} \geq 0$,
we obtain from Eq.~(\ref{eq:lower-bound}) that
$\tan{\gamma_-} \geq L_- = 1.12$,
meaning that $\gamma_- \geq 48^\circ$ is guaranteed with a rather
mild theoretical assumption.
This lower bound on $\gamma_-$ can be seen clearly in
FIG.~\ref{figure1}.

Unfortunately,
we must contend with the discrete ambiguities.
First we notice that,
due to the twofold discrete ambiguity in the inversion
of the function $\tan{\gamma_-}$,
$L_-$ also produces the bound $\gamma_- \geq -132^\circ$,
for $\gamma_-$ in the range $-180^\circ \leq \gamma_- < 0^\circ$.
We can exclude solutions with negative $\gamma$ if we assume that
there is no new physics in the $K - \overline{K}$ system
(and trust the sign of the relevant hadronic matrix element).

We must also consider the bound from $L_+$.
Since both $C \cot{\delta}$ and $I_+$ are positive,
we obtain $\tan{\gamma_+} \geq L_+$.
This means that $\gamma_+ \geq 174^\circ$,
for $\gamma \in (0^\circ, 180^\circ)$;
or $\gamma_+ \geq -6^\circ$,
if we take $\gamma \in (-180^\circ, 0^\circ)$.
In both ranges of $\gamma$,
the bound from $L_+$ is much tighter than the bound from
$L_-$.
We conclude that $\gamma$ is constrained by the bound from
$L_-$.

We can see from FIG.~\ref{figure1} that our assumption
of positive $C \cot{\delta} \geq 0$ plays a crucial role.
Indeed,
when we cross $C \cot{\delta} = 0$ the lower bounds become upper
bounds.
Moreover,
in the region of negative $C \cot{\delta}$,
$\gamma_+$ goes through a region of vary rapid variation
and it even crosses $\gamma_-$.
This occurs for 
\begin{equation}
C \cot{\delta} =
- \frac{S + \sin{(2 \tilde{\beta})}}{\cos{(2 \tilde{\beta})}}
= - 0.223,
\end{equation}
as can be seen in the figure and understood from
Eq.~(\ref{eq:crossing}).
The usual assumption that $C \cot{\delta}$
(which is proportional to $\cos{\delta}$) is positive,
hinges on the belief that the magnitude of $\delta$ should be small
and that the corresponding matrix element should have the sign
obtained from factorization.
However,
it could be that the ratio of ``penguin to tree'' has
a sign opposite to that taken from factorization,
in which case $\delta \sim 180^\circ$ and $C \cot{\delta}$
would be negative \cite{Wolf-sign}.
We have shown that,
if that is the case,
this analysis can still be performed,
but with the lower bounds becomming upper bounds.
Unfortunately in this case $L_+$ will provide the effective upper bound
$\gamma \leq 174^\circ$, which is useless.
It is important to stress that,
for $\gamma$ negative,
the assumption consistent with QCD factorization is 
$C \cot{\delta}<0$,
as is evident from the $\sin{\gamma}$ term in 
Eq.~(\ref{eq:C-from-theor}) and from the previous
discussion.

\begin{figure}[htb]
\includegraphics*[height=7cm]{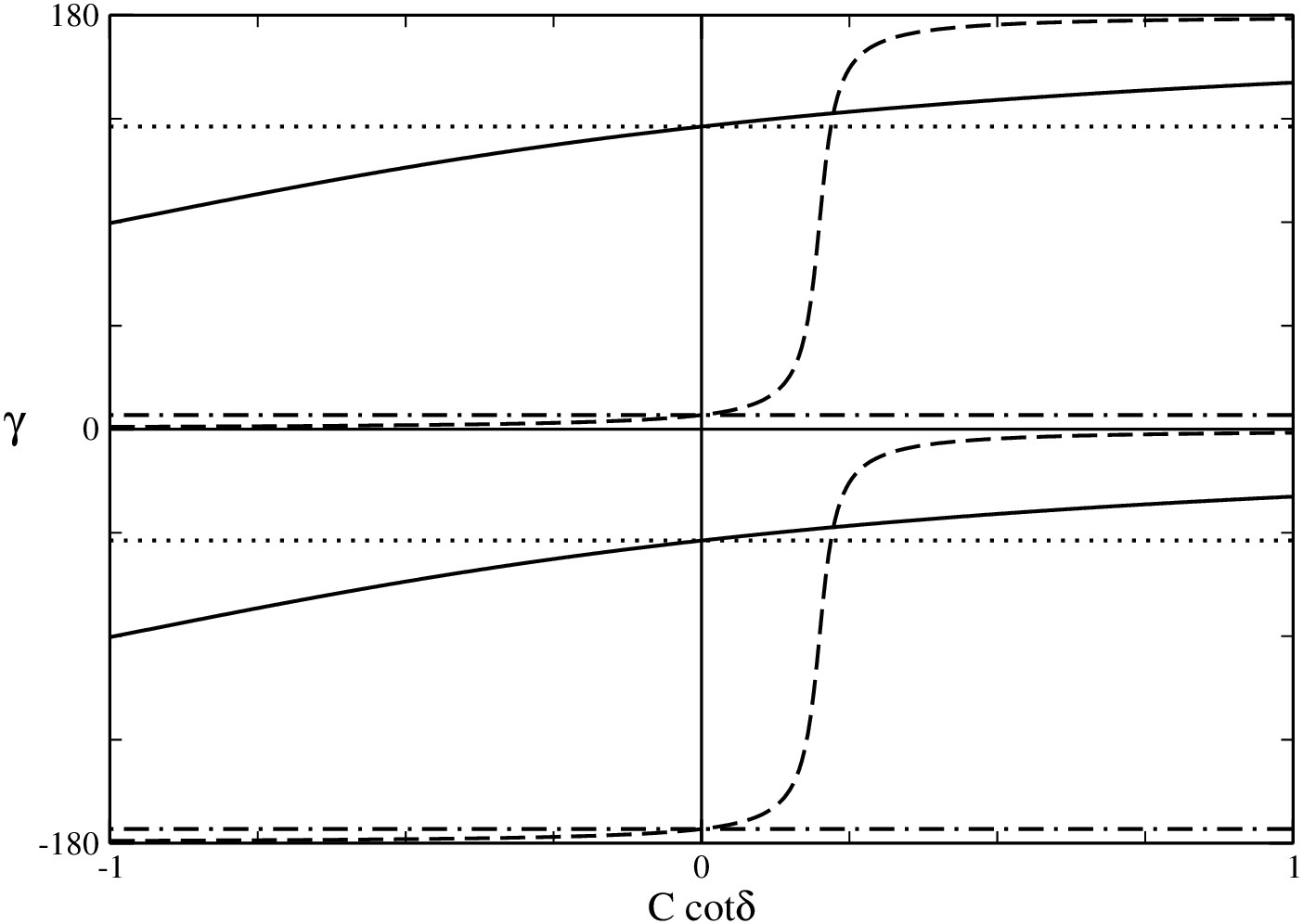}
\caption{\label{figure2}Values ``determined'' for $\gamma$,
as a function of the theoretical input for $C \cot{\delta}$,
assuming $\sin{(2 \tilde{\beta})}= 0.736$,
$C=0.0528$,
and $S=-0.585$.
The solid (dashed) curves correspond to $\gamma_+$ ($\gamma_-$).
The horizontal dotted (dash-dotted) lines correspond to
the values of $\gamma$ for which $\tan{\gamma}=L_+$
($\tan{\gamma}=L_-$).
Here we choose $\cos{(2 \tilde{\beta})}$ negative.
}
\end{figure}

The problem of $\gamma_+$ crossing $\gamma_-$,
seen in FIG.~\ref{figure1} for $C \cot{\delta} < 0$,
will come back to haunt us when we consider the
possibility that $\cos{(2 \tilde{\beta})} < 0$,
because of the
$\cos{(2 \tilde{\beta})} \rightarrow -\cos{(2 \tilde{\beta})}$,
$\cot{\delta} \rightarrow - \cot{\delta}$,
$\tan{\gamma_\pm} \rightarrow - \tan{\gamma_\mp}$
symmetry we alluded to at the end of section~\ref{sec:master}.
This symmetry is clear from the comparizon of
FIG.~\ref{figure1} with
FIG.~\ref{figure2},
drawn for the same 
$\sin{(2 \tilde{\beta})}= 0.736$,
$C=0.0528$,
and $S=-0.585$ ``experimental'' observables,
but assuming the negative
$\cos{(2 \tilde{\beta})} = - 0.677$ possibility.

For $\cos{(2 \tilde{\beta})} = - 0.677$,
we obtain
$I_+=0.992$, $L_+ = -1.12$, $I_- = -0.195$ and $L_- = 0.11$.
(This was to be expected from the fact that
$\cos{(2 \tilde{\beta})} \rightarrow - \cos{(2 \tilde{\beta})}$
leads to $L_\pm \rightarrow - L_\mp$.)
If we keep our assumption that $C \cot{\delta} \geq 0$,
then $\tan{\gamma_+} \geq L_+$,
meaning that $\gamma_+ \geq 132^\circ$,
if we take $\gamma \in (0^\circ, 180^\circ)$,
or $\gamma_+ \geq -48^\circ$,
if we take $\gamma \in (-180^\circ, 0^\circ)$.
Again,
we may assume the SM in the $K - \overline{K}$ system to
exclude the last possibility.

Unfortunately,
$L_-$ only provides the very poor bound
$\arctan{L_-} = 6^\circ$.
It is true that this problem can be avoided by
ignoring the $\cos{(2 \tilde{\beta})} < 0$ solution.
But,
if we are assuming new physics,
we should not discard this possibility in an ad-hoc way
(as is sometimes done).
Amusingly,
when $\cos{(2 \tilde{\beta})} < 0$,
it is the assumption that factorization yields the wrong
sign for $\delta$
(and, thus, that $C \cot{\delta} < 0$)
that provides us with bounds on $\gamma$ in the
$(0^\circ, 180^\circ)$ region.

\begin{figure}[htb]
\includegraphics*[height=7cm]{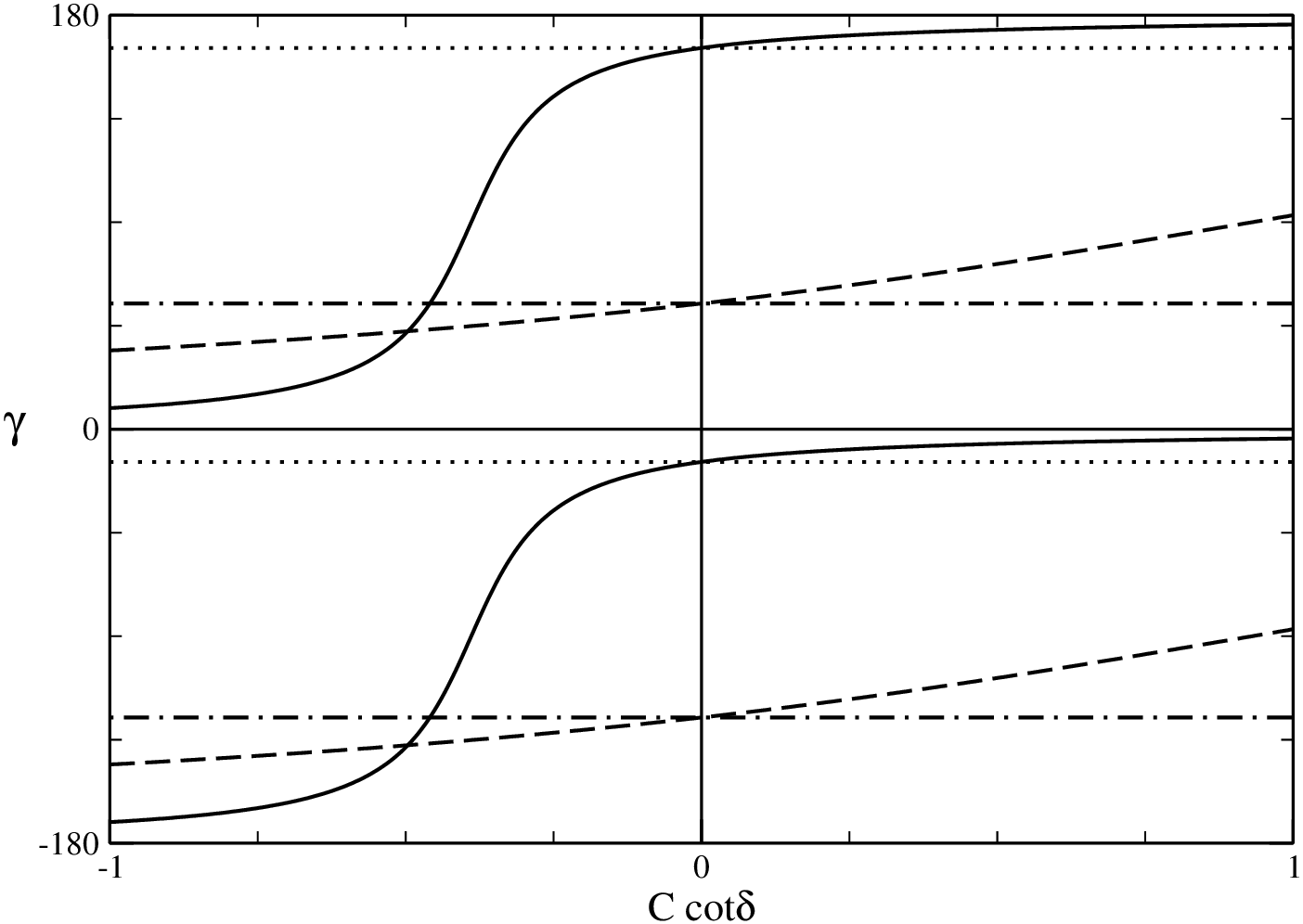}
\caption{\label{figure3}Values ``determined'' for $\gamma$,
as a function of the theoretical input for $C \cot{\delta}$,
assuming $\sin{(2 \tilde{\beta})}= 0.736$,
$C=-0.19$,
and $S=-0.40$.
The solid (dashed) curves correspond to $\gamma_+$ ($\gamma_-$).
The horizontal dotted (dash-dotted) lines correspond to
the values of $\gamma$ for which $\tan{\gamma}=L_+$
($\tan{\gamma}=L_-$).
Here we choose $\cos{(2 \tilde{\beta})}$ positive,
as in the SM.
}
\end{figure}

The previous case was motivated by the theoretical expectations
$r\, R_b=0.11$, $\delta=0.15$, and $\gamma=60^\circ$.
FIG.~\ref{figure3} shows the same analysis performed for
$\sin{(2 \tilde{\beta})}= 0.736$,
$\cos{(2 \tilde{\beta})}$ positive,
and assuming the current Babar central values
$C=-0.19$,
and $S=-0.40$ \cite{Babar-pipi}.
The solution with $\cos{(2 \tilde{\beta})}$ negative
can be obtained through the symmetry already described.
We find that $\gamma \geq 55^\circ$,
if we take $C \cot{\delta} \geq 0$
and $\gamma \in (0^\circ, 180^\circ)$.

\subsection{\label{sec:BS}Recovering the Buchalla-Safir bound}

Buchalla and Safir \cite{BS} have considered a particular case
of the bounds in Eq.~(\ref{master-Joao}),
which corresponds to setting $C=0$ and assuming the SM.
Indeed,
they argue that the current (SM) constraints on $\gamma$
eliminate all solutions except the one arising from
the canonical inversion of $L_-$.
Indeed,
restricting $\gamma$ to $(0, \pi)$ eliminates four solutions;
assuming the SM eliminates the possibility that
$\cos{(2 \tilde{\beta})} < 0$;
and the other current bounds on $\gamma$ eliminate $\gamma_+ $.

Taking
$C \cot{\delta}\ (C \cot{\delta} + I_-)\ I_- >0$,
we obtain
\begin{equation}
\tan{\gamma_-} \geq L_- \, .  
\end{equation}
This would be the simple generalization of the BS bound
to the $C \neq 0$ case,
if we were to ignore all the discrete ambiguities.

Their (lowest) bound is obtained by setting $C$ to zero
in Eq.~(\ref{eq:L+-}):
\begin{eqnarray}
L_-^0
& = &
\frac{-1 - \sqrt{1 - S^2} \cos{(2 \beta)}
- S \sin{(2 \beta)}}{S \cos{(2 \beta)}
- \sqrt{1 - S^2} \sin{(2 \beta)}},
\label{eq:L0-Joao}
\\
& = &
\frac{- \cos{(2 \beta)} - \sqrt{1 - S^2} + 1
+ S \sin{(2 \beta)} + \cos{(2 \beta)} \sqrt{1 - S^2}
}{
S - \sin{(2 \beta)} - S \cos{(2 \beta)}
+ \sqrt{1 - S^2} \sin{(2 \beta)}},
\label{eq:L0-BS}
\\
& = &
\frac{\cos{(2 \beta)}
+ \sqrt{1 - S^2}
}{
\sin{(2 \beta)} - S}.
\label{eq:L0-Quico}
\end{eqnarray}
Eq.~(\ref{eq:L0-Joao}) results directly from Eq.~(\ref{eq:L+-});
Eq.~(\ref{eq:L0-BS}) is a slightly rewritten version of the
expression in \cite{BS}.
They are both equal to the simplest form in Eq.~(\ref{eq:L0-Quico}).
Now,
\begin{equation}
L_- - L_-^0
=
\frac{(\sqrt{1-S^2} - \sqrt{1-C^2 -S^2}) (S + \sin{(2 \beta)})}{
\left[S \cos{(2 \beta)} - \sqrt{1 - C^2 - S^2} \sin{(2 \beta)}\right]
\left[S \cos{(2 \beta)} - \sqrt{1 - S^2} \sin{(2 \beta)}\right]
}.
\label{90again}
\end{equation}
Since $\sqrt{1 - C^2 - S^2} \leq \sqrt{1 - S^2}$,
the numerator is positive whenever $S \geq - \sin{(2 \beta)}$.
It is true that the denominator will be negative when the two terms
between the square brackets have opposite signs.
But that only occurs because $\arctan{L_-}$ goes through
$90^\circ$ before $\arctan{L_-^0}$,
and it does not affect the order of the bounds on $\gamma$ \cite{explain}.
As a result,
eliminating all the discrete ambiguities,
we recover the BS bound
\begin{equation}
\tan{\gamma_-} \geq L_-^0,
\end{equation}
which is valid for $S \geq - \sin{(2 \beta)}$ \cite{BS}.

\begin{figure}[htb]
\includegraphics*[height=7cm]{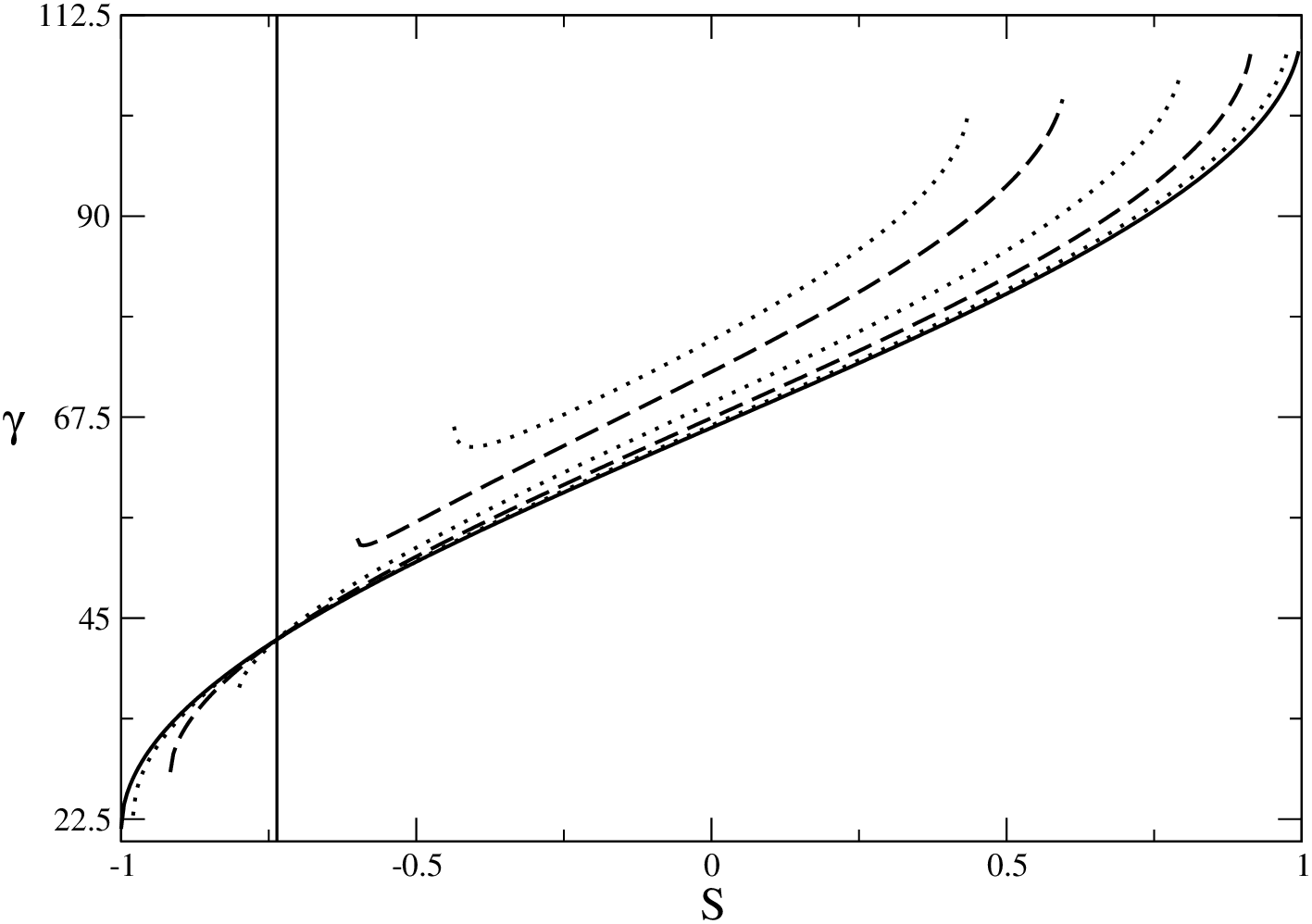}
\caption{\label{figure4}Bounds on $\gamma$ obtained
from $L_-$ as a function of $S$,
for different choices of $|C|$.
The vertical line corresponds to $S = - \sin{(2 \beta)}$.
To the right of it,
the solid line (which corresponds to $C=0$) lies 
below all other lines,
in accordance with the BS bound.
The other lines correspond to
$|C|=0.2$,
$|C|=0.4$,
$|C|=0.6$,
$|C|=0.8$,
and $|C|=0.9$,
going from the bottom up.
}
\end{figure}

FIG.~\ref{figure4} compares the bounds on $\gamma$ obtained
from $L_-$ as a function of $S$,
for different choices of $|C|$
(notice that the value of $L_-$ does \textit{not} depend on
the sign of $C$).
We have taken $\sin{(2 \tilde{\beta})}= 0.736$
and $\cos{(2 \tilde{\beta})} = + 0.677$.
The vertical line corresponds to $S = - \sin{(2 \beta)}$.
To the right of it,
the solid line (which corresponds to $C=0$) lies 
below all other lines,
in accordance with the BS bound.
The dotted line immediately above was obtained with $|C|=0.2$
and the others correspond to $|C|=0.4$,
$|C|=0.6$,
$|C|=0.8$,
and $|C|=0.9$,
respectively.
FIG.~\ref{figure4} shows that our $L_-$ bound improves on the BS bound,
and that its impact becomes more relevant for large $|C|$ values.

We can now understand why Buchalla and Safir required the
constraint $S \geq - \sin{(2 \beta)}$.
They did so for two reasons.
First, because the BS bound ($C=0$) only lies
below the lines with $C \neq 0$ in that case,
as seen clearly in FIG.~\ref{figure4}.
Second because,
for $\gamma \in (0^\circ, 180^\circ)$,
$L_+$ lies below $L_-$ if $S < - \sin{(2 \beta)}$.
This can be understood from Eq.~(\ref{eq:crossing}),
recalling that the expressions for $L_\pm$ are obtained
from those of $\tan{\gamma_\pm}$ by setting
$C \cot{\delta}$ to zero.
But this means that there in nothing fundamental about
Buchalla and Safir's restriction that
$S \geq - \sin{(2 \beta)}$.
Indeed,
for $S \geq - \sin{(2 \beta)}$, 
the bound from $L_-$ with $C=0$ provides the
(lowest) lower bound on $\gamma$.
But,
for $S < - \sin{(2 \beta)}$, 
the bound from $L_-$ with $C=0$ is still useful since it
provides the (highest) upper bound on $\gamma$.

\subsection{\label{subsection:Q}Dependence of the analysis on the
theoretical parameter $Q$}

In the previous sections we have used $C \cot{\delta}$ as the
one piece of theoretical input required to extract $\gamma$
from the $B \rightarrow \pi^+ \pi^-$ decays.
This was choosen in order to compare our new constraints
based on $L_\pm$ (valid for any $C$ and for $\tilde{\beta} \neq \beta$)
with that obtained by Buchalla and Safir in the limit $C=0$.
However,
the quantity $Q = r \cos{\delta}$ is easier to constrain
theoretically and also allows the extraction of $\gamma$.
Indeed, one can show that Eq.~(\ref{master-Quico}) yields
\begin{equation}
\gamma_\pm =
- \arctan{\left[ \frac{1-R_\pm}{I_\pm} \right]}
+ \arcsin{\left[ \frac{Q\, (R_\pm -1)}{
\sqrt{I_\pm^2 + (1 - R_\pm)^2}} \right]}.
\label{eq:Quico2}
\end{equation}
For the ``-'' sign,
the first term reproduces the BS bound,
while the second term shows the correction for $Q \neq 0$.

Given a set of experimental values for
$\sin{(2 \tilde{\beta})}$, $S$, and $C$,
the theoretical parameter $Q$ cannot take any value
at random.
Fortunately,
the limits that those experiments place on $Q$
are built into Eq.~(\ref{eq:Quico2}),
since the magnitude of the argument of the function
$\arcsin{}$ cannot exceed unity.
Therefore
\begin{equation}
Q^2
\leq
\frac{I_\pm^2 + (1 - R_\pm)^2}{(1 - R_\pm)^2}
=
\frac{2(1 - R_\pm) - C^2}{(1 - R_\pm)^2}.
\label{eq:bound-on-Q}
\end{equation}

\begin{figure}[htb]
\includegraphics*[height=7cm]{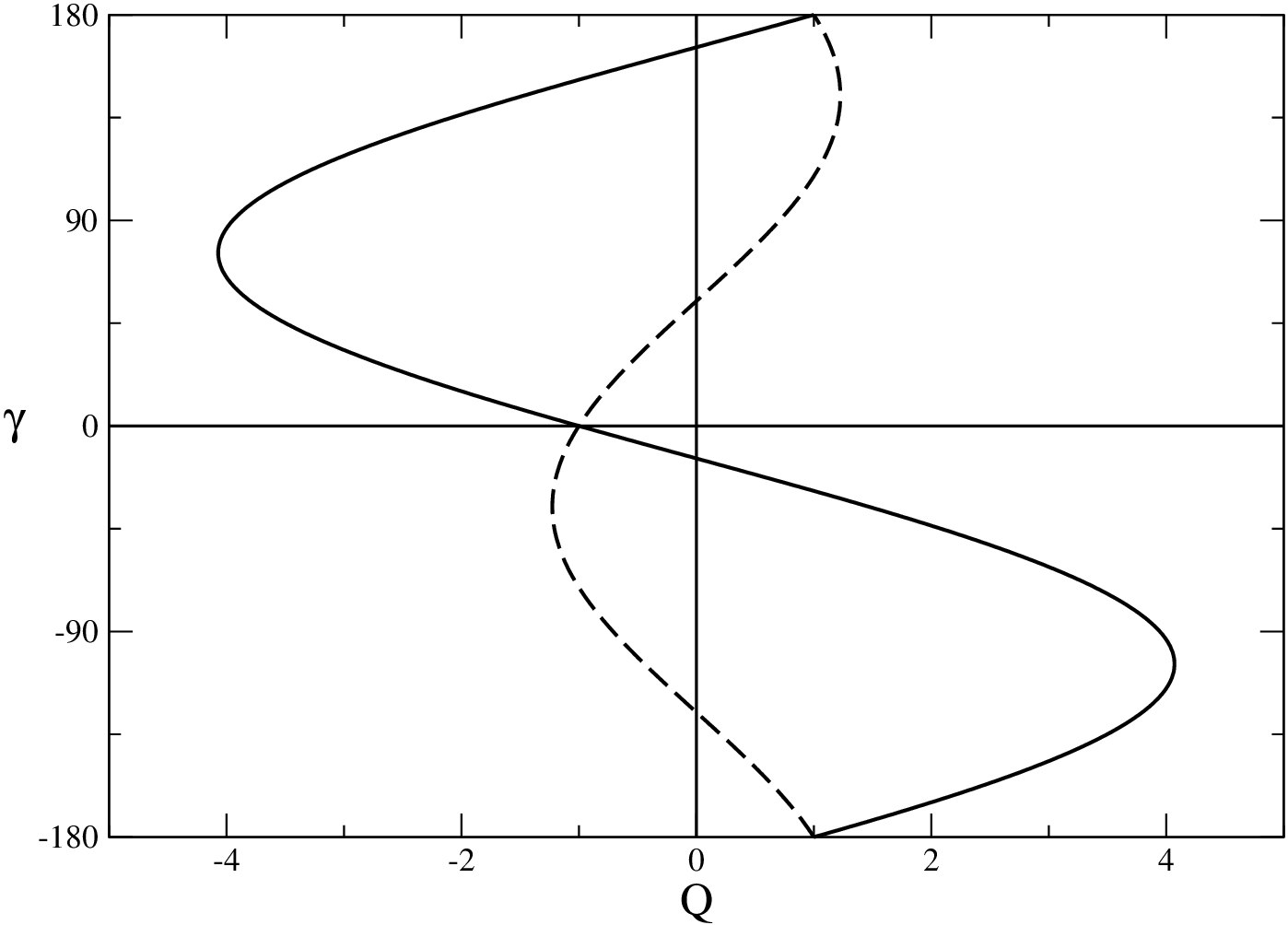}
\caption{\label{figure5}Values ``determined'' for $\gamma$,
as a function of the theoretical input for $Q$,
assuming $\sin{(2 \tilde{\beta})}= 0.736$,
$C=-0.19$,
and $S=-0.40$.
The solid (dashed) curve corresponds to $\gamma_+$ ($\gamma_-$).
Here we choose $\cos{(2 \tilde{\beta})}$ positive,
as in the SM.
}
\end{figure}

FIG.~\ref{figure5} shows the dependence of $\gamma_\pm$
on $Q$ for the same ``experimental'' values used
in FIG.~\ref{figure3}.
Namely,
$\sin{(2 \tilde{\beta})}= 0.736$,
$\cos{(2 \tilde{\beta})}$ positive,
and assuming the current Babar central values
$C=-0.19$,
and $S=-0.40$ \cite{Babar-pipi}.
The solution with $\cos{(2 \tilde{\beta})}$ negative
is easily obtained through the transformations
$\gamma_\pm \rightarrow - \gamma_\mp$,
as can be seen directly in Eq.~(\ref{master-Quico}).
As in FIG.~\ref{figure3},
we find that $\gamma \geq 55^\circ$,
if we take $Q \geq 0$
and $\gamma \in (0^\circ, 180^\circ)$.

Incidentally,
Eq.~(\ref{eq:bound-on-Q}) gives for $|Q|$ the upper bounds of
$1.2$ and $4.1$,
for the $\gamma_-$ and $\gamma_+$ branches, respectively,
in accordance with FIG.~\ref{figure5}.

\section{\label{sec:conclusions}Conclusions}

The extraction of the CKM angle $\gamma$ from the time-dependent
decay $B \rightarrow \pi^+ \pi^-$ requires one piece of external input.
Here we have studied the depedence of that analysis on
the theoretical parameters $C \cot{\delta}$ or $r \cos{\delta}$.
Of course,
a similar analysis can be performed with any other input information,
such as experimental input from the isospin analysis \cite{GL}
or the SU(3) relation with $B \rightarrow K^+ \pi^-$ \cite{SW94}.
The novelty introduced by Buchalla and Safir,
is that a mild assumption about the theoretical parameters
already allows interesting bounds to be placed on $\gamma$ \cite{BS}.

We have extended their result in several ways:
i) we have provided a simpler derivation of their bound,
which avoids the Wolfenstein parameters $\rho$ and $\eta$;
ii) we have pointed out that the restriction to
$S \geq - \sin{(2 \tilde{\beta})}$ is not fundamental in the sense
that, for $S < - \sin{(2 \tilde{\beta})}$, the only change
is that the ``lowest'' bound becomes a ``highest'' bound;
iii) we have highlighted the impact that new physics phases in the
$B - \overline{B}$ mixing have, discussing, in particular,
the possibility that $\cos{(2 \tilde{\beta})}$ might be negative;
iv) we have extended their bounds to the case $C \neq 0$.
Naturally,
the methods applied here can be used in other decays \cite{us}.

\begin{acknowledgments}
We are grateful to G.\ C.\ Branco and L.\ Wolfenstein
for discussions and to L.\ Lavoura for carefully
reading and commenting on this manuscript.
This work was supported by the Portuguese \textit{Funda\c{c}\~{a}o para
a Ci\^{e}ncia e a Tecnologia} (FCT) under the contract CFIF-Plurianual (2/91).
In addition, F.\ J.\ B.\ is partially supported by by FCT under
POCTI/FIS/36288/2000 and by the spanish M.\ E.\ C.\ under FPA2002-00612,
and J.\ P.\ S.\ is supported by FCT under POCTI/FNU/37449/2001.
\end{acknowledgments}

\appendix*
\section{\label{appendix}A useful inequality}
Consider the function
\begin{equation}
f(\theta) = A \sin{\theta} + B \cos{\theta}.
\end{equation}
Its derivative is zero when
\begin{equation}
\sin{\theta} = \frac{\pm A}{\sqrt{A^2+B^2}},
\ \ \ \ \ \ \ \ 
\cos{\theta} = \frac{\pm B}{\sqrt{A^2+B^2}}.
\end{equation}
At these points, $f(\theta)$ takes the extremum values
$\pm \sqrt{A^2+B^2}$. Therefore,
\begin{equation}
-\sqrt{A^2+B^2} \leq A \sin{\theta} + B \cos{\theta} \leq \sqrt{A^2+B^2}.
\label{eq:luis}
\end{equation}


\begin{thebibliography}{99}
\bibitem{Dun95}
This quantity would be measureable (and this discrete ambiguity removed)
if the $B - \overline{B}$ lifetime difference were experimentaly
measurable.
This may occur for the $B_s - \overline{B_s}$ system,
where the point was first made by
I.\ Dunietz,
Phys.\ Rev.\ D \textbf{52}, 3048 (1995).
\bibitem{note2}
This may be taken directly from page 116
of Ref.~\cite{BLS},
with the substitutions:
$\phi_M \rightarrow - \tilde{\beta}$,
$\phi_1 \rightarrow \gamma + \tilde{\beta}$,
$\phi_2 \rightarrow \tilde{\beta}$,
$\Delta \rightarrow \delta$,
and $\eta_f \rightarrow 1$.
One must also change the sign of $\lambda$ 
due to the different convention for $q/p$
used there.
This is a good place to point out one other subtlety:
strictly speaking,
it is only after one chooses the convention
$\Delta m_B > 0$
(as is implicitly done in most calculations of
$q/p$)
that the sign of $\lambda$ acquires physical significance.
\bibitem{explain}
This is clear from FIG.~\ref{figure4}.
Consider for example $S=0.5$.
There,
the curve of $\gamma$ for $C=0.8$ has already
gone through $90^\circ$,
while the BS curve ($C=0$) has not.
As a result,
the denominator of Eq.~(\ref{90again}) is negative and
$L_- \leq L_-^0$.
Nevertheless,
the bound from the $C=0.8$ line lies above the bound
from the $C=0$ line.
This is a trivial artefact resulting from the inversion
of the function $\tan{(x)}$,
and it has no physical meaning.
\end{thebibliography}
\end{document}